\documentclass[12pt]{iopart}
\usepackage{iopams}
\usepackage{graphicx}
\usepackage{caption}
\usepackage{subcaption}
\usepackage{hyperref}
\usepackage{hypernat}
\usepackage{color}

\begin{document}

\title[Random walks on networks with preferential cumulative damage]{Random walks on networks with preferential cumulative damage: Generation of bias and aging}
\author{L. K. Eraso-Hernandez${}^1$, A. P. Riascos${}^{2}$, T.M. Michelitsch${}^{3}$ and J. Wang-Michelitsch${}^4$}  
\address{${}^1$Departamento de F\'isica, Universidad de los Andes, A.A. 4976, Bogot\'a D.C., Colombia\\
${}^2$Instituto de F\'isica, Universidad Nacional Aut\'onoma de M\'exico, 
Apartado Postal 20-364, 01000 Ciudad de M\'exico, M\'exico\\
${}^3$Sorbonne Universit\'e, Institut Jean le Rond d'Alembert, CNRS UMR 7190,4 place Jussieu, 75252 Paris cedex 05, France\\
${}^4$Independent researcher, Paris, France}


\begin{abstract}
In this paper, we explore the reduction of functionality in a complex system
as a consequence of cumulative random damage and imperfect reparation, a phenomenon modeled as a dynamical process on networks. We analyze the global characteristics of the diffusive movement of random walkers on networks where the walkers hop considering the capacity of transport of each link. The links are susceptible to damage that generates bias and aging. We describe the algorithm for the generation of damage and the bias in the transport producing complex eigenvalues of the transition matrix that defines the random walker for different types of graphs, including regular, deterministic, and random networks. The evolution of the asymmetry of the transport is quantified with local information in the links and further with non-local information associated with the transport on a global scale such as the matrix of the mean first passage times and the fractional Laplacian matrix. Our findings suggest that systems with greater complexity live longer.
\end{abstract}
%
%
\newpage
\section{Introduction}
The study of dynamical processes taking place on networks has a significant impact in different fields of science and engineering, leading to important applications in diverse contexts \cite{VespiBook}. In particular, the dynamics of a random walker that hops visiting the nodes of the network following different strategies is an important issue due to connections with a vast field of interdisciplinary topics like the ranking of the Internet \cite{GooglePR1998}, transport on networks \cite{LambiottePRE2011}, the modeling of human mobility in urban settlements \cite{RiascosMateosPlos2017}, chemical reactions \cite{Chemical}, digital image processing \cite{GradyRW2006}, algorithms for extracting useful information from data \cite{Tremblay2014,FoussBook2016},  epidemic spreading \cite{SatorrasPRL2001,Barter_PRE2016,bestehorn2020markovian}, and the huge field of continuous-time random walks including unbiased and biased anomalous transport and fractional dynamics  \cite{MontrollWeiss1965,MetzlerKlafter2000,GorenfloMainardi2008,KutnerMasoliver2017,Barkai2001,WangBarkai2020} (and many others)   
just to mention a few examples related to the study of complex systems.
\\[2mm]
One of the main features observed in many ``complex systems'' at different scales such as
living organisms, complex materials, social systems, companies, and civilizations, is that these systems
exhibit aging and a limited lifespan associated with damage impact and degradation of the functionality. In many cases, these systems have to maintain a global function (for example transport) and in order to ``survive'', the system needs continuously respond to damage impact with ``reparation processes'' maintaining both immediate and long-term survival. The reparation process in a complex system is performed under a time constraint that the functionality of the entire system during the reparation has to be maintained. Due to this time constraint of the reparation process, when the damage impact is ``too severe'',  the system is not able to re-establish the original undamaged structure but generates a repaired structure with altered properties a so-called ``{\it misrepair}'' \cite{WangMiWun2009}. This mechanism can be considered as a compromise between two needs: Reparation as good as possible and as fast as necessary. In many cases, the alteration in a misrepaired structure compared to the initial undamaged structure may be very `small' and the misrepair may be even `close' to perfect repair. Therefore, the mechanism of misrepair guarantees immediate survival as a result of a reparation process; however, the price to be paid for this fast reparation process is a misrepaired altered structure with reduced functionality compared to the initial undamaged structure \cite{WangMi2015}.
\\[2mm]
Despite of its importance for understanding aging as a dynamical process in complex systems, the modeling of aging associated with the accumulation of damage (`misrepairs') has been less explored. Recent advances in this direction support the idea that aging occurs as an emergent phenomenon in many biological structures and systems \cite{MitnitskiPRE2016,Mitnitski2017,Aging_PhysRevE2019,DuanIEEE2020,LedbergPlosOne2020,SunPNAS2020}. In this contribution, we study the aging process of a complex system as a consequence of a preferential accumulation of damage. We model this system as a network with weights that evolve in time due to damage generating bias on the links and the global transport of this structure. In this case, aging is a consequence of the reduction of the system's capacity to communicate within the whole structure, modeled by ergodic random walks that navigate through the links of the network. In the initial configuration of the system, we assume the network to be a ``perfect structure'' described by a connected network gradually altered by progressing accumulation of damage. We analyze the evolution of asymmetry and aging for networks with different topologies allowing us to distinguish their complexities qualitatively by considering their connectivity and robustness to maintain a specific function.
\section{Transport on networks with cumulative damage}
\label{SectionModel}
In this section, we present a phenomenological model for aging processes in a complex system represented
by a network. The model relies on three characteristics \cite{Aging_PhysRevE2019}: 1) We consider a network for which the nodes and
connections contribute collectively to its global functionality which includes transport processes \cite{Hughes,MasudaPhysRep2017},
synchronization \cite{Arenas2008}, diffusion \cite{BlanchardBook2011,FractionalBook2019}, among others \cite{VespiBook}.  
The capacity of this structure to perform these functions is measured by a global quantity in a determined configuration of the system. 2) The entire system is subjected to stochastic damage that reduces the functionality of the links affecting the global activity, this detriment is cumulative.  3) We compare the global functionality of the system with the initial state (with optimal conditions and no damage). The definition of a threshold value for the functionality
required for the system to operate allows determining if the system is alive.
\\[2mm]
The features 1)-3) may exist in different complex systems. The gradual deterioration of the global functionality resulting from the accumulation of misrepairs in a complex system is the subject of the model to be developed and explored in the present section. The model incorporates the observed phenomena of self-amplification in the occurrence of misrepairs, i.e. a structure that is already altered by misrepairs is more likely to `attract' further misrepairs \cite{WangMi2015b}. We also account for the observation that there are two temporal scales. One is the ``fast'' time scale of functional operation; for example, the temporal evolution of a random walk. The second ``slow'' time scale is where the dynamics of accumulation of misrepairs and aging take place. These assumptions reflect the observed fact that metabolic functions in a living organism may take some seconds to a few hours, whereas aging changes in living beings may take years.
\subsection{Network structure and cumulative damage}
For the initial configuration, we consider undirected connected networks with $N$ nodes $i=1,\ldots ,N$ described by an adjacency matrix $\mathbf{A}$ with elements $A_{ij}=A_{ji}=1$ if there is an edge between the nodes $i$ and $j$ and $A_{ij}=0$ otherwise; in particular, $A_{ii}=0$ to avoid edges connecting a node with itself. In this structure, we denote the set of nodes as $\mathcal{V}$ and the set of edges (links) as $\mathcal{E}$ with elements $(i,j)$. For each pair in $\mathcal{E}$, the corresponding element of the adjacency matrix is non-null. In the following the link from $i$ to $j$ denoted as $(i,j)$ is independent from $(j,i)$, and we denote as $|\mathcal{E}|$ the total number of different edges in the network.
\\[2mm]
Additionally to the network structure, the global state of the system at time $T=0,1,2,\ldots$ is characterized by a $N\times N$ matrix of weights $\mathbf{\Omega}(T)$ with elements $\Omega_{ij}(T)\geq 0$ and $\Omega_{ii}(T)=0$ which describe weighted connections between the nodes. The matrix $\mathbf{\Omega}(T)$ contains information of the state of the edges and in general is not symmetric. In order to capture in the model the damage impact affecting the complex system, we introduce the variable $T$ as the measure of the total number of damage hits in the links of the network. The variable $T$ 
can also be conceived as a time measure by assuming a constant damage impact rate, i.e. successive damage events occur with a constant difference of times $\Delta T=1$. We introduce for each link $(i,j) \in \mathcal{E}$ a stochastic integer variable $h_{ij}(T)$ where $h_{ij}(T)-1$ counts the number of
random faults that exist in this link at time $T$. The values $h_{ij}(T)$ for all the edges are numbers that evolve randomly, and a new fault in the link $(i,j)$ appears at time $T$ with a probability $\pi_{ij}(T)$ which is given by
\begin{equation}\label{problinks}
\pi_{ij}(T)=\frac{h_{ij}(T-1)}{\sum_{(l,m) \in \mathcal{E}} h_{lm}(T-1)}\qquad (i,j) \in \mathcal{E},
\end{equation}
for $\,T=1,2,\ldots$ with the initial condition $h_{ij}(0)=1$, i.e. no faults exist for all the edges at $T=0$. Equation (\ref{problinks}) indicates the probability for the event that at time $T$ the number of faults $h_{ij}(T)=h_{ij}(T-1)+1$ are increased by one.
In our analysis, the damage is distributed without maintaining the symmetry of the initial undirected network. In this manner, in the general case $h_{ij}(T)$ is independent of the value $h_{ji}(T)$ and also $\pi_{ij}(T)$ from $\pi_{ji}(T)$ which is generating a biased network. 
With Eq. (\ref{problinks}) at $T=1$ the first hit (fault) is randomly generated for any selected link $(i,j)$
with equal probability $\pi_{ij}(1)=\frac{1}{|\mathcal{E}|}$. The occurrence of the second fault at $T=2$ depends on the previous configuration and so on. An asymptotic analysis of the time-evolution of the fault number distribution resulting from Eq. (\ref{problinks}) 
shows that a power-law scaling with features of a stochastic fractal emerge (See Eq. (\ref{result}) in \ref{Append1} and consult also Ref. \cite{Aging_PhysRevE2019}).
An essential feature of the probabilities in Eq. (\ref{problinks}) is that they produce preferential damage if a link has already suffered damage in the past. A link has a higher probability to get a fault with respect to a link never being damaged. Such preferential random processes have been explored in different contexts in science (see Ref. \cite{NetworkScienceBook2016}), being a key element in our model that generates complexity in the distribution of damage reflected by asymptotically emerging power-law and fractal features (See \ref{Append1}).
\\[2mm]
The choice of generating law of faults in Eq. (\ref{problinks}) is based on the observation of aging changes in living beings. The development of aging changes such as age spots is self-amplifying and inhomogeneous. Age spots develop in this way because misrepaired structures have increased damage-sensitivity and reduced reparation-efficiency \cite{WangMi2015,WangMi2015b}.
\\[3mm]
Now, we aim to describe how the structure reacts to the damage hits occurring stochastically to the edges. We describe the effects of the damage by using the information in the matrix of weights $\mathbf{\Omega}(T)$. In terms of the values $h_{ij}(T)$, the matrix $\mathbf{\Omega}(T)$ defines the global state of the 
network containing the complete information on the network topology at time $T$.
Its matrix elements 
\begin{equation}\label{OmegaijT}
\Omega_{ij}(T)=(h_{ij}(T))^{-\alpha} A_{ij}
\end{equation}
contain the local information on the damaged state of edge $(i,j)$
and $\alpha\geq 0$ is a real-valued parameter that quantifies the effect of the damage in each link. We call $\alpha$ the {\it misrepair parameter} since it describes the capacity of the system to repair damage in the links: In the limit $\alpha\to 0$ the system responds with perfect reparation with $\Omega_{ij}(T) \to A_{ij}$ as in a perfect undamaged structure, and the effect of the stochastically generated faults is null. In contrast, in the limit $\alpha\to\infty$, a hit in a link is equivalent to its removal from the network. This limit corresponds to a complex system without repair capacity. The fault accumulation dynamics described by Eq. (\ref{problinks}) together with the `misrepair equation' (\ref{OmegaijT}) is therefore able to mimic the phenomena related to aging processes observed in living beings \cite{WangMiWun2009,Kirkwood2005}.
\begin{figure}[!t] 
\begin{center}
\includegraphics*[width=1.0\textwidth]{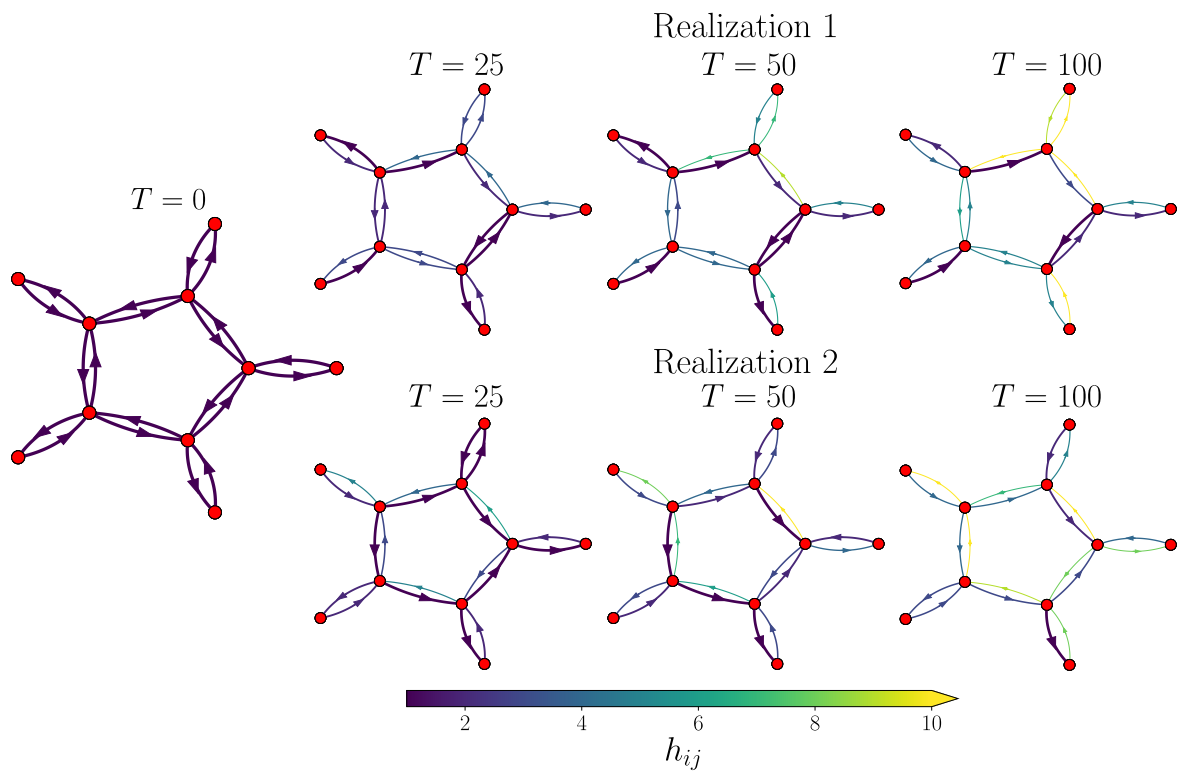}
\end{center}
\vspace{-5mm}
\caption{\label{Fig1}(Color online) Monte Carlo simulation of the reduction of functionality in a network with random faults in the links. We generate a random hit in the edges at times $T=1,2,\ldots$ and in each line, we have a value $h_{ij}$ that depends on $T$. At each time $T$, the value $h_{ij}-1$ gives the number of random hits that the link $(i,j)$ has suffered (considering the initial values $h_{ij}=1$ at $T=0$). The probability to have a new fault in one of the edges is determined by the Eq. (\ref{problinks}). By using this algorithm,  we implement Monte Carlo simulations to generate random faults in the graph described at time $T=0$. The values in the colorbar indicate $h_{ij}$ for all the links,  we use the value $\alpha=2$ in Eq. (\ref{OmegaijT}). We present the configurations of the system at times $T=0,\, 25,\, 50,\, 100$ for two different realizations of this process. } 
\end{figure}
In Fig. \ref{Fig1}, we illustrate the algorithm for the generation of cumulative damage in a network with $N=10$ nodes formed by a comb-like structure with a ring. For this directed graph, we generate random hits in the links at times $T=1,2,\ldots,100$. The probability to generate a hit at time $T$ in the directed edge $(i,j)$ is proportional to the previous configuration given by $h_{ij}(T-1)$ in Eq. (\ref{problinks}). In this way, using Monte Carlo simulations, we depict the damage in two realizations.
The values of $h_{ij}$ in the edges are represented with colors
codified in the colorbar
whereas the respective widths reflect
their capacity of transport given by $\Omega_{ij}(T)$ in Eq. (\ref{OmegaijT})
with the parameter $\alpha=2$. The results in the final configuration at $T=100$ give us an idea of the variety 
of structures how the damage can be distributed in the network affecting the nodes in the periphery or in the central ring.
\subsection{Master equation and mean first passage times}

Let us now come back  to the general algorithm for cumulative damage described before, at a completely different scale of times (significantly less than the characteristic times for the damage) takes place the movement of a random 
walker in the network with discrete steps $\Delta t$ at times $t=0,\Delta t, 2\Delta t,\ldots$.  
For each state of the network's weights, 
the random walker can move visiting nodes and 
is capable to visit all the nodes of the structure that changes very slowly with time $T$. 
In a determined configuration at time $T$, the transition probability matrix $\mathbf{W}(T)$ describing the random walker is defined by the elements $w_{i\to j}(T)$ with the probability to pass from node $i$ to node $j$
\begin{equation}\label{transitionPij}
w_{i\to j}(T)=\frac{\Omega_{ij}(T)}{\sum_{\ell=1}^N\Omega_{i\ell}(T)}.
\end{equation}
We assume a {\it Markovian} 
time-discrete random walker that performs at any time increment $\Delta t$ a random step from one node to another.
This process is defined by the master equation \cite{LambiottePRE2011,Hughes,NohRieger2004}
\begin{equation}
 \label{mastereqnNRW}
 P_{ij}(t+\Delta t,T) = \sum_{\ell=1}^NP_{i\ell}(t,T)w_{\ell\to j}(T) 
\end{equation}
valid for $t \ll \Delta T =1$. In this master equation $P_{ij}(t,T)$ indicates the probability that the walker that starts its walk at node $i$ at $t=0$ occupies node $j$ at the $n$-th time step $t=n\Delta t$.
\\[2mm]
The one-step transition matrix in Eq. (\ref{transitionPij}) is constructed such that the walker has to change the node at any step
(i.e $w_{i\to i}(T)=0$).
The canonical representation of the ($t=n\Delta t$) of the $n$-step transition matrix is
\begin{equation}
 {\mathbf{P}}(n\Delta t,T) = ({\mathbf W}(T))^n=
\sum_{m=1}^N (\lambda_m(T))^n |\phi_m(T)\rangle\langle {\bar \phi}_m(T)|.
 \label{mastereqnNRWtimeevo}
\end{equation}
Where we use Dirac's (bra-ket) notation. In Eq. (\ref{mastereqnNRWtimeevo}), $|\phi_m(T)\rangle, \langle {\bar \phi}_m(T)|$ denote, respectively, the right and left  eigenvectors of the transition matrix with the respective eigenvalues $0\leq |\lambda_m(T)| \leq 1$. The walk which we assume to take place on a (strongly connected) directed weighted and finite network is ergodic with 
the unique eigenvalue $\lambda_1(T) =1 \forall T $ reflecting row stochasticity (with corresponding right-eigenvector having identical components)
of the transition matrix $\sum_{j=1}^N w_{i \to j}(T) = 1 $ and $|\lambda_m(T)| \leq 1$ maintained $ \forall T$ for $m=1,\ldots ,N$ (see Ref. \cite{FractionalBook2019} for a detailed analysis of undirected networks, and Refs. \cite{DirectedFractional_PRE2020,tmm-fp-apr_fractal-fract-2020} for directed networks).
The stationary distribution $ P_{j}^{(\infty)}(T)\equiv \lim_{t\to\infty}\frac{1}{t}\sum_{t^\prime=0}^tP_{ij}(t^\prime,T) $, that gives the probability to find the random walker at the node $j$ in the limit $t\to\infty$, is given by \cite{MasudaPhysRep2017,FractionalBook2019,RiascosMateos2012}
\begin{equation}\label{stationary}
P_{j}^{(\infty)}(T) = \langle i|\phi_1(T)\rangle\langle \bar \phi_1(T)|j\rangle
\end{equation}
where, since $\left\langle i|\phi_1(T)\right\rangle=\mathrm{constant}$ (independent of $i$), the stationary distribution $P_j^{(\infty)}(T)$ does not depend on the initial condition. 
\\[2mm]
Additionally, we have the mean first passage time $\left\langle {\cal T}_{ij}\right\rangle $ that gives the average number of time steps (in units of $\Delta t$) the walker needs 
to travel from node $i$ to node $j$ in the form (see Refs.  \cite{FractionalBook2019,DirectedFractional_PRE2020,RiascosMateos2012,ZhangPRE2013} for a complete derivation)
\begin{equation}
\label{TijSpect}
\hspace{-10mm}
\left\langle {\cal T}_{ij}(T)\right\rangle = \frac{1}{P_{j}^{(\infty)}(T) }\left[\delta_{ij}+\sum_{\ell=2}^N \frac{\left\langle j|\phi_\ell(T)\right\rangle \left\langle\bar{\phi}_\ell(T)|j\right\rangle-\left
 \langle i|\phi_\ell(T)\right\rangle \left\langle\bar{\phi}_\ell(T)|j\right\rangle}{1-\lambda_\ell(T)} 
 \right].
\end{equation}
In this relation for $i=j$ the second term and for $j\neq i$ the first term vanishes.
For $i=j$ this relation gives
the {\it mean first return time} or {\it mean recurrence time} \cite{FractionalBook2019,Kac1947}
\begin{equation}
\label{meanrcurrencetime}
\left\langle {\cal T}_{jj}(T)\right\rangle =\frac{1}{{\left\langle j|\phi_1(T)\right\rangle \left\langle\bar{\phi}_1(T)|j\right\rangle}}.
\end{equation}
This is the {\it Kac-formula} relating the mean recurrence time with the inverse of the
stationary distribution $P_{j}^{(\infty)}(T)$. 
\\[2mm]
\begin{figure}[!t] 
 \centering
\includegraphics*[width=1.0\textwidth]{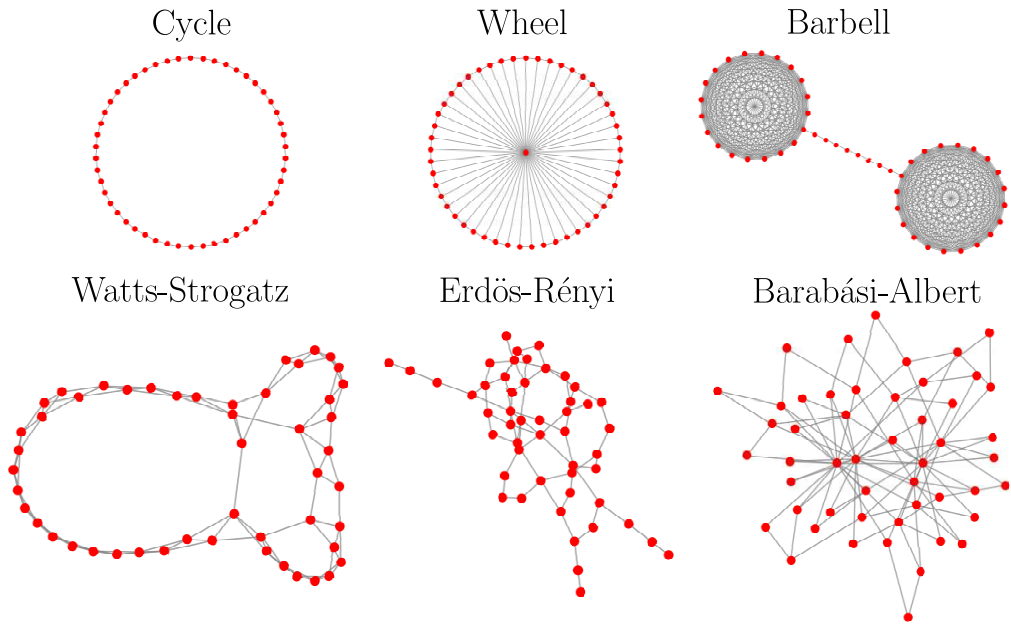}  
\caption{\label{Fig2} Different types of networks with $N=50$ nodes. }
 \end{figure}
\begin{figure}[!t] 
\begin{center}
\includegraphics*[width=0.9\textwidth]{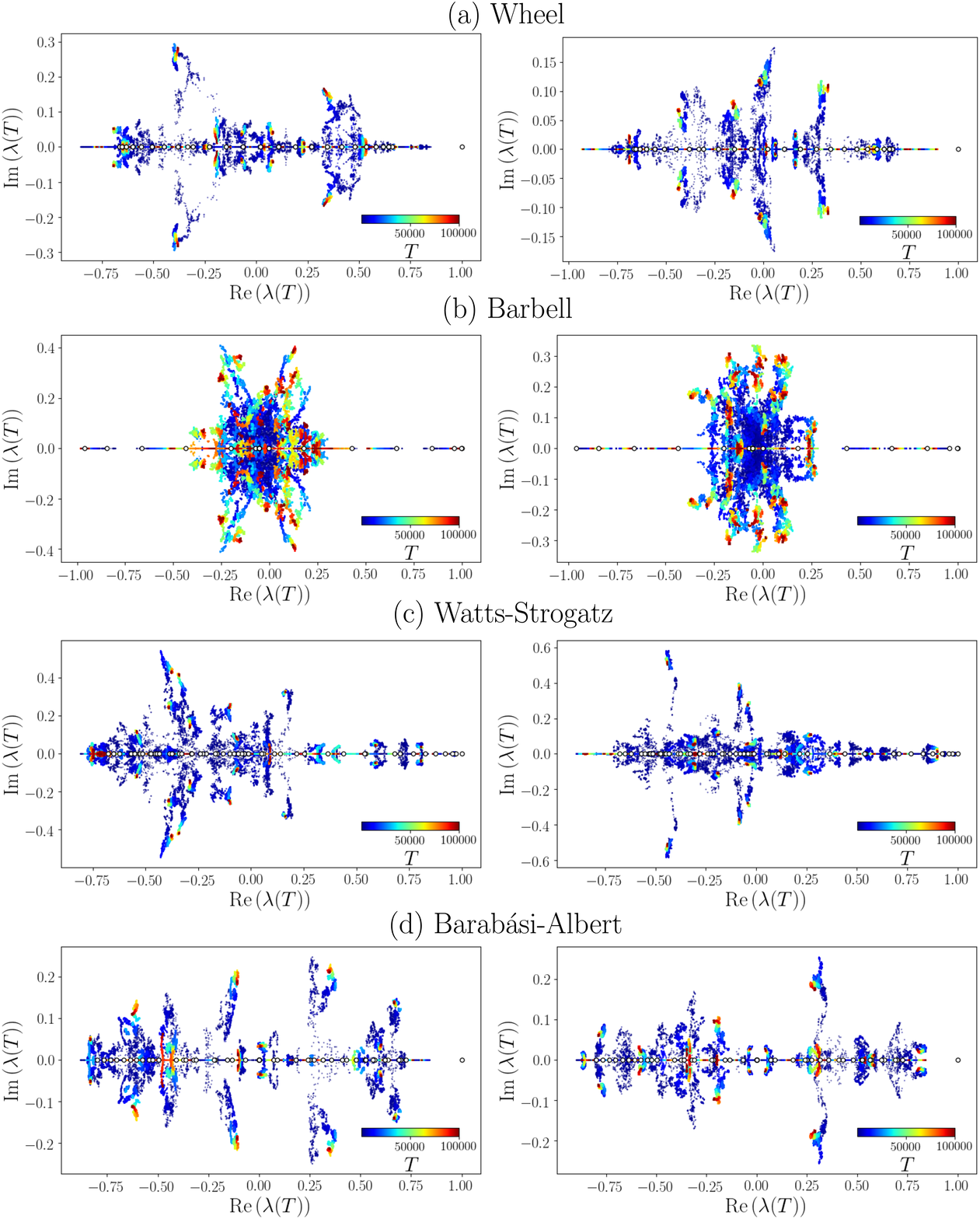}
\end{center}
\vspace{1mm}
\caption{\label{Fig3} Eigenvalues of the transition matrices $\mathbf{W}(T)$ for networks with $N=50$ nodes with cumulative damage in the edges. We show the results for the set of eigenvalues $\lambda(T)$ in the complex plane for different times $T=0,1,\ldots,10^5$ codified in the colorbar. The initial structures without damage are the undirected networks: (a) wheel graph, (b) barbell graph, (c) random network generated with the Watts-Strogatz algorithm, and (d)  a Barab\'asi-Albert random network. For each case we depict two realizations of the cumulative damage with $\alpha=1$ in Eq. (\ref{OmegaijT}), we represent with circles the eigenvalues of the transition matrix without damage $\mathbf{W}(0)$.} 
\end{figure}
The results in Eqs. (\ref{mastereqnNRWtimeevo}) and (\ref{TijSpect}) show that the eigenvalues of the transition matrix $\mathbf{W}(T)$ at different times $T$ contain important information of the diffusive transport of each configuration at times $T=0,1,\ldots$. Also, due to the asymmetry in some edges, in the general case, the eigenvalues $\lambda_j(T)$ can take complex values. However, the process of cumulative damage for $\alpha$ finite maintains the ergodicity of the random walker described by $\mathbf{W}(T)$ since the initial structure is connected and none of the edges is completely removed with the hits at different times (for a detailed discussion about ergodicity on directed networks see Ref. \cite{DirectedFractional_PRE2020} and references therein) thus the mean first passage times in Eqs. (\ref{TijSpect})-(\ref{meanrcurrencetime}) are well defined. The relation between the eigenvalues of the transition matrix and node or edge removal strategies in various dynamical processes on networks has been explored numerically by Restrepo et. al. in Refs. \cite{Restrepo_Ott_Hunt_2006,restropo2008prl}.
\\[2mm]
In the following part, we explore the effects of cumulative damage in the network topologies illustrated in Fig. \ref{Fig2}. In Fig. \ref{Fig3} we depict the eigenvalues of the transition matrix $\mathbf{W}(T)$ for different initial networks with $N=50$ nodes and two realizations of the random cumulative damage algorithm. Figure \ref{Fig3}(a) displays the results for a wheel graph formed by a single node connected to all nodes of a cycle (with $49$ nodes). In Fig. \ref{Fig3}(b) we show the eigenvalues for a barbell graph, {\it i.e.}, a network with two well-defined communities composed of two fully connected networks (with $20$ nodes each) connected by a linear graph (with $10$ nodes) \cite{Ghosh2008Barbell}. Conversely, in Figs. \ref{Fig3}(c)-(d) the initial structures without damage are random networks obtained with the Watts-Strogatz algorithm \cite{WattsStrogatz1998} generated from a ring with nearest-neighbor and next-nearest-neighbor links and a rewiring probability of $p=0.05$ and the Barab\'asi-Albert algorithm, generated with the preferential attachment rule \cite{BarabasiAlbert1999}.
\\[2mm]
The results in Fig. \ref{Fig3} show the evolution of the eigenvalues with the damage at times $T=1,2,\ldots,100000$, the initial eigenvalues are real and are represented with circles in each case. The damage in each directed edge induce asymmetry and the bias generated produces complex eigenvalues depicted with different colors. Here it is worth mentioning that we apply the same approach to several types of trees where we found that in this type of graph the eigenvalues are real. This is an important case for which we do not have an analytical result but we hope this leads to future research in the spectral properties of directed weighted trees. These findings motivate the introduction of a measure to quantify the evolution of bias generated with the accumulation of damage in different networks. We discuss this point in the following part.
\section{Quantifying the evolution of bias }
In this section we discuss different ways to quantify the asymmetry in the diffusive transport generated by the systematic accumulation of damage in the links. We introduce local measures considering the elements of the transition matrix $\mathbf{W}(T)$ and non-local quantities that include global information of the dynamics.
\subsection{Local information}
\subsubsection{$\mathcal{S}_W(T)$.} The first measure implemented to characterize the asymmetry of the diffusive transport at time $T$ is the direct comparison of the elements $(i,j)$ and $(j,i)$ of the transition matrix $\mathbf{W}(T)$. We use the quantity  
 \begin{equation}\label{SWcomp}
  \mathcal{S}_W(T)=1-\frac{2}{N(N-1)-2M}\sum_{i=1}^N\sum_{j=i+1}^N\frac{|w_{i\to j}(T)-w_{j\to i}(T)|}{w_{i\to j}(T)+w_{j\to i}(T)}, 
  \end{equation}
 introduced in Ref. \cite{esposito_measuring_2014} to quantify the asymmetry in the connectivity of neural networks. In Eq. (\ref{SWcomp}) $w_{i\to j}(T)$ represents the elements of the $N\times N$ transition matrix $\mathbf{W}(T)$ and $M$ is the number of times for which both $w_{i\to j}(T)$ and $w_{j\to i}(T)$ are zero, these cases are not included in the sums in Eq. (\ref{SWcomp}).  If  $\mathcal{S}_W(T)$ is close to $1$, the probabilities $w_{i\to j}(T)$ and $w_{j\to i}(T)$ are close and the random walker has  similar probabilities to go forward and backward in each connection. Conversely, a value close to $0$ shows  that those weights and probabilities are quite different. Therefore, this parameter allows us to quantify the local bias of the transition matrix in each configuration of the process at times $T=0,1,2,\ldots$. 
 \\[2mm]
 \subsubsection{Random walk entropy.} The asymmetry of the transport can be measured indirectly from an information perspective. In this way, the information to define the random walk strategy in the structure with damage evolves with $T$. We use the local information \cite{Weng_Small_Zhang_Hui_2015}
 \begin{equation}\label{Entropy}
  \mathcal{H}(T)=-\sum_{i,j=1}^N P_i^{(\infty)}(T) w_{i\to j}(T)\log\left[w_{i\to j}(T)\right],
 \end{equation}
 where $P_i^{(\infty)}(T)\equiv \langle i|\phi_1(T)\rangle\langle \bar{\phi}_1(T)|i\rangle$ is the $i$-th component of the stationary distribution. In this way, we define the measure $ \mathcal{S}_{\mathrm{Entropy}}(T)$ as
 \begin{equation}\label{S_Entropy}
  \mathcal{S}_{\mathrm{Entropy}}(T)=\frac{\mathcal{H}(T)}{\mathcal{H}(0)}
 \end{equation}
 that gives the ratio between entropy $\mathcal{H}(T)$ at time $T$ with the initial value  $\mathcal{H}(0)$ calculated from the structure without damage.
\\[2mm]
The value $\mathcal{H}(T)$ measures the minimum amount of information necessary to carry out the diffusion process in the network and depends on the topology and dynamics \cite{Gomez-Gardenes_Latora_2008}. Through the maximization of the entropy rate, it is possible to design optimal diffusion processes and define random walkers that are maximally dispersing, which means that they can perform every possible walk with the same probability in the graph. Normal random walks on regular lattices are a particular case of maximum entropy rate, since their nodes have the same degree all the trajectories of a given length are equiprobable \cite{Sinatra_PRE2011}, \cite{Burda_Duda_Luck_Waclaw_2009}.  Thus, it makes sense to propose the entropy rate as a measure of asymmetry, since aging is changing the probabilities of certain paths in the graph, also as the damage increases, some paths are less probable than others so the information to describe the random walk strategy decreases with $T$ and it is more difficult for the random walker to reach all nodes in the network. In our definition  in Eq. (\ref{S_Entropy}) we are interested in the evolution of $\mathcal{S}_{\mathrm{Entropy}}(T)$ with the initial condition $\mathcal{S}_{\mathrm{Entropy}}(0)=1$. 
\subsection{Global information}
\subsubsection{$\mathcal{S}_{\mathrm{MFPT}}$.} The local measure $\mathcal{S}_W(T)$ in Eq. (\ref{SWcomp}) only considers the elements of the transition matrix; however, relevant  information of the process is included in all the paths connecting two nodes on the network. In this way we modify Eq. (\ref{SWcomp}) to define
 \begin{equation}\label{SMFPT}
  \mathcal{S}_{\mathrm{MFPT}}(T)=1-\frac{2}{N(N-1)}\sum_{i=1}^N\sum_{j=i+1}^N\frac{|\langle {\cal T}_{ij}(T)\rangle-\langle {\cal T}_{ji}(T)\rangle|}{\langle {\cal T}_{ij}(T)\rangle+\langle {\cal T}_{ji}(T)\rangle}, 
 \end{equation}
where the elements $w_{i\to j}(T)$ were replaced by the mean first passage time $\langle {\cal T}_{ij}(T)\rangle$ in Eq. (\ref{TijSpect}), that gives the average number
of steps that a discrete-time random walker defined by $\mathbf{W}(T)$ needs to move from node
$i$ and  reach $j$ for the first time. In contrast with the definition in Eq. (\ref{SWcomp}), now we use the value $M=0$ since from Eq. (\ref{TijSpect}), $\langle {\cal T}_{ij}(T)\rangle>0$ as a consequence of the ergodicity of the network at any finite $T$.
\subsubsection{$\mathcal{S}_{\gamma}$.}
A different way to have access to non-local information in diffusive transport is through the fractional Laplacian $\mathbf{L}^\gamma$ of a graph \cite{FractionalBook2019,TMM_APR_recurrence2017,RiascosMateosFD2014,Perkins2019}, a formalism  explored for directed networks in Refs. \cite{DirectedFractional_PRE2020,benzi2020fractional}. In terms of the matrix of weights $\Omega_{ij}(T)>0$, the Laplacian matrix $\mathbf{L}$ with elements $i$, $j$ is given by \cite{DirectedFractional_PRE2020}
\begin{equation}\label{Laplacian}
L_{ij}(T)=k_i^{(\mathrm{out})}(T) \delta_{ij}-\Omega_{ij}(T)
\end{equation}
with the out-degree $k_i^{(\mathrm{out})}(T) =\sum_{\ell=1}^N\Omega_{i\ell}(T)$. The fractional Laplacian $\mathbf{L}^\gamma$ with $0<\gamma<1$ satisfies the following conditions \cite{FractionalBook2019,DirectedFractional_PRE2020}: {\bf (i)} For the {\it fractional out-degree}, we have
\begin{equation}
\label{fracout}
 k_i^{(\gamma)}\equiv(\mathbf{L}^\gamma)_{ii}=
 -\sum_{m\neq i} (\mathbf{L}^\gamma)_{im}.   
\end{equation}
{\bf (ii)} The diagonal elements of $\mathbf{L}^\gamma$ are positive real values; in this way  $k_i^{(\gamma)}>0$ for $i=1,2,\ldots, N$ and, {\bf (iii)} The non-diagonal elements of $\mathbf{L}^\gamma$ are real values
satisfying $(\mathbf{L}^\gamma)_{ij} < 0$ for $i\neq j$ ($\gamma \in (0,1)$) in ergodic networks
(and $(\mathbf{L}^\gamma)_{ij} \leq 0$ in non-ergodic cases, see Ref. \cite{DirectedFractional_PRE2020} for a discussion of ergodicity in directed networks).
\\[2mm]
The characteristics of the fractional Laplacian matrix allow to define the fractional diffusion on directed weighted networks as a discrete-time Markovian process determined by a transition matrix $\mathbf{W}^{(\gamma)}(T)$ with elements $w_{i\to j}^{(\gamma)}(T)$ representing the probability to hop from $i$ to $j$ given by \cite{DirectedFractional_PRE2020}
\begin{equation}\label{wijfrac}
w_{i\to j}^{(\gamma)}(T)=\delta_{ij}-\frac{(\mathbf {L}^\gamma)_{ij}(T)}{(\mathbf{L}^\gamma)_{ii}(T)}\qquad 0<\gamma\leq 1,
\end{equation}
these transition probabilities combine the information of all possible paths connecting  nodes $i$ and $j$ on the network \cite{DirectedFractional_PRE2020}. We can use this property to quantify the asymmetry of the transport generated with the introduction of damage; in this way, we apply Eq. (\ref{SWcomp}) for  $\mathbf{W}^{(\gamma)}(T)$ 
 \begin{equation}\label{Sgamma}
  \mathcal{S}_\gamma(T)=1-\frac{2}{N(N-1)}\sum_{i=1}^N\sum_{j=i+1}^N\frac{|w_{i\to j}^{(\gamma)}(T)-w_{j\to i}^{(\gamma)}(T)|}{w_{i\to j}^{(\gamma)}(T)+w_{j\to i}^{(\gamma)}(T)}
  \end{equation}
 valid for $0<\gamma<1$. In the case with $\gamma=1$, the fractional dynamics approach recovers the normal random walk with local displacements; however, Eq. (\ref{Sgamma}) only applies for 
 $0<\gamma<1$, where for this interval in ergodic networks all the non-diagonal elements of $\mathbf{W}^{(\gamma)}(T)$ 
 are non-null \cite{DirectedFractional_PRE2020,tmm-fp-apr_fractal-fract-2020}.
\subsection{Asymmetry and cumulative damage}
\begin{figure}[!t]
	\centering
	\includegraphics*[width=\textwidth]{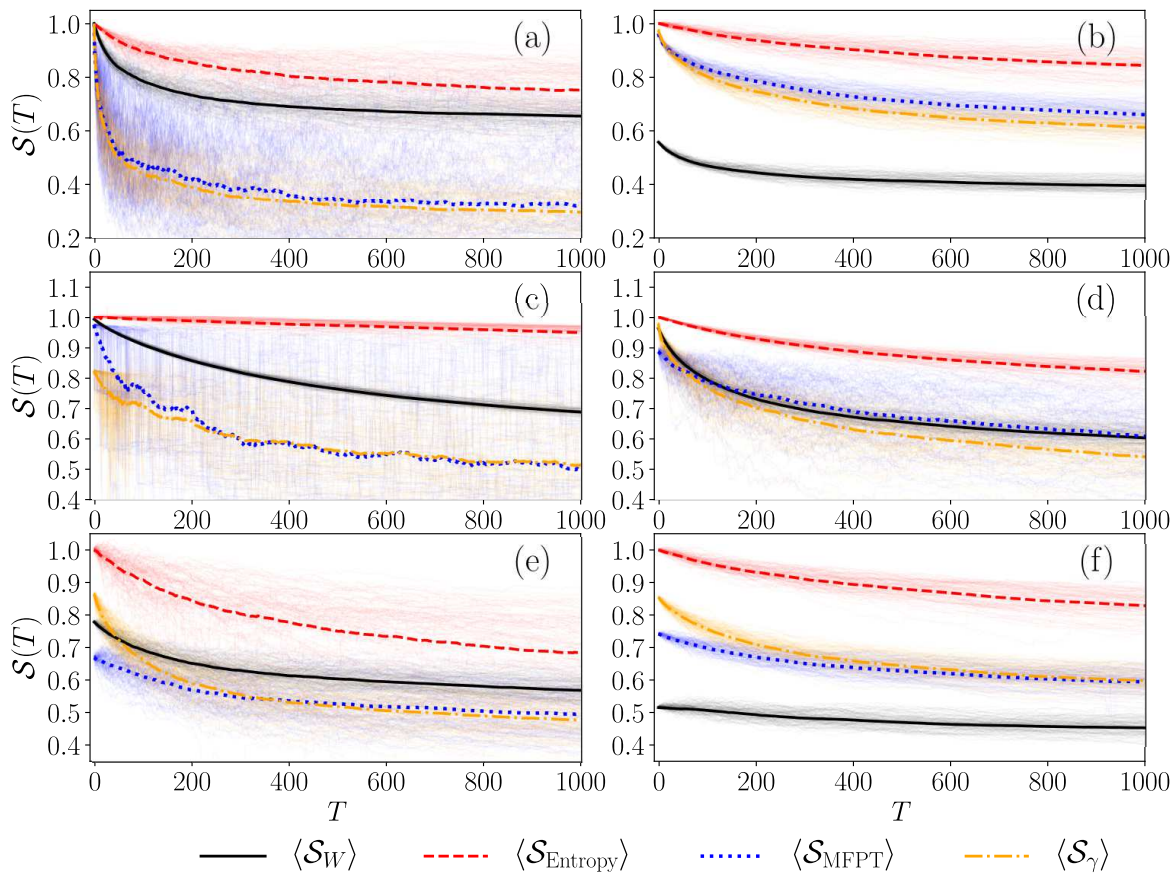}    
	\caption{\label{Fig4} Average evolution of asymmetry generated by cumulative damage in networks with $N=50$. We present the results for the value $S(T)$  as a function of $T$ for three deterministic networks: (a) a ring, (b) a wheel graph, (c) a barbell graph, and three random networks:  (d) a Watts-Strogatz network with rewiring probability $p=0.05$, (e) an Erd\"{o}s-R\'enyi (ER) network at the connectivity threshold $p=\log(N)/N$, and (f)  a Barab\'asi-Albert (BA) network. In this analysis $S(T)$ denotes the quantities $\mathcal{S}_W(T)$, $\mathcal{S}_{\mathrm{Entropy}}(T)$, $\mathcal{S}_{\mathrm{MFPT}}(T)$, and $\mathcal{S}_\gamma(T)$ with $\gamma=0.5$, we present with thin lines the results obtained for $100$ realizations of the  cumulative damage algorithm with $\alpha=1$ and their respective ensemble average. }
\end{figure}
\begin{figure}[!t] 
 \centering
 \includegraphics*[width=\textwidth]{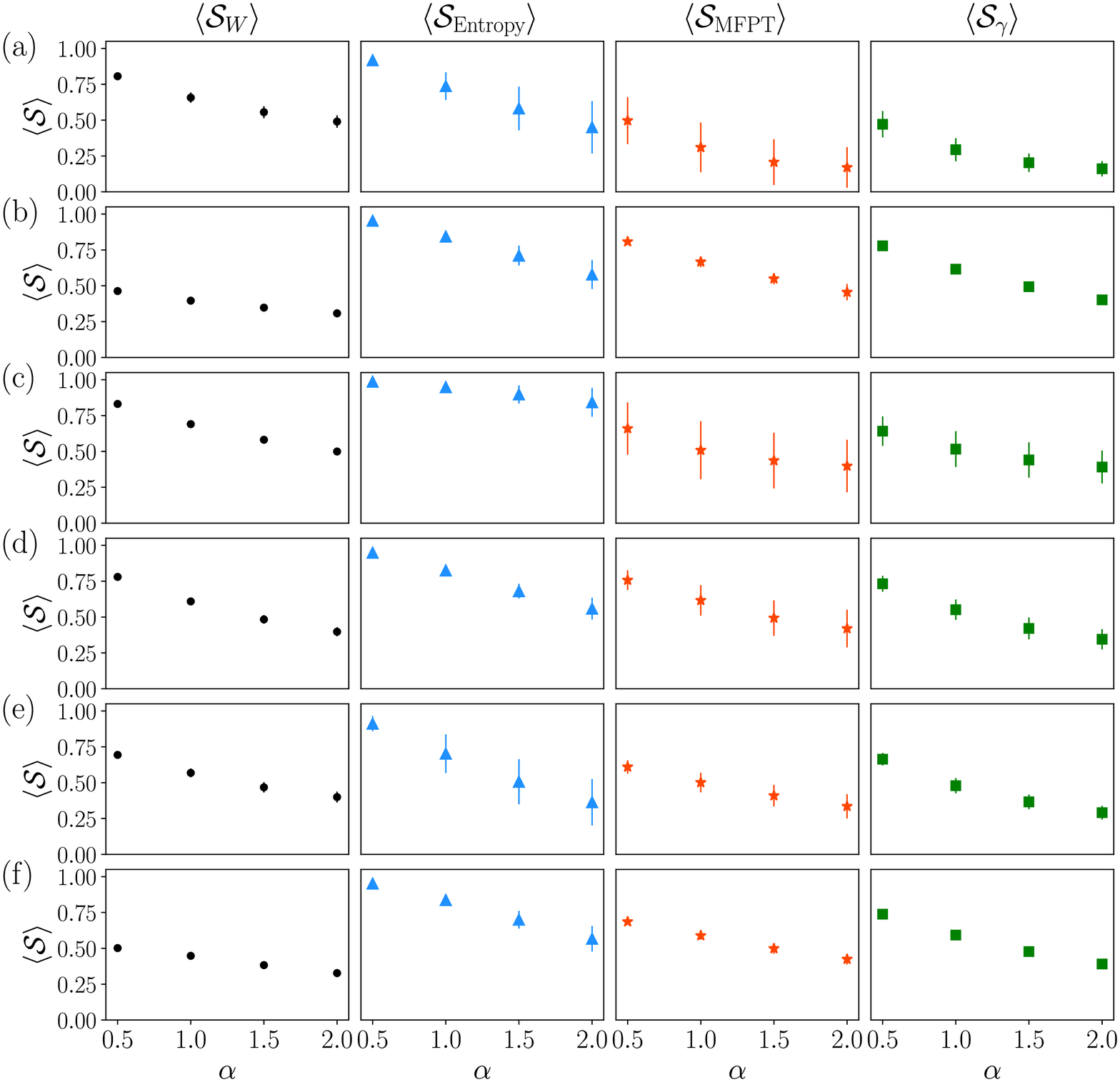}    
\caption{\label{Fig5} Dependence of the ensemble average of asymmetry measures $\left\langle\mathcal{S}\right\rangle$ for different values of the parameter $\alpha$. In (a)-(f) we apply the algorithm of cumulative damage for the networks analyzed in Fig. \ref{Fig4}. The results show the ensemble average of $\mathcal{S}_W(T)$, $\mathcal{S}_{\mathrm{Entropy}}(T)$, $\mathcal{S}_{\mathrm{MFPT}}(T)$, and $\mathcal{S}_\gamma(T)$ with $\gamma=0.5$ at time $T=1000$ considering 1000 realizations for $\alpha=0.5,1,1.5,2$, error bars denote the standard deviation of the values. }
 \end{figure}

Once defined four measures to quantify the asymmetry on transport on networks, we explore the evolution of these quantities with time $T$ for structures with $N=50$ nodes. We analyze three deterministic networks: a ring, a wheel and a barbell graph, and three random structures generated with the Watts-Strogatz (WS), Erd\"{o}s-R\'enyi (ER), Barab\'asi-Albert (BA) algorithms. In Fig. \ref{Fig4} we depict the evolution of $\mathcal{S}(T)$ that denotes $\mathcal{S}_W(T)$, $\mathcal{S}_{\mathrm{Entropy}}(T)$, $\mathcal{S}_{\mathrm{MFPT}}(T)$, and $\mathcal{S}_\gamma(T)$ considering a fractional non-locality with $\gamma=0.5$. The results were obtained from Monte Carlo simulations of the accumulation of damage algorithm with a misrepair parameter $\alpha=1$ in Eq. (\ref{OmegaijT}). For each time $T$ we calculate the values of $\mathcal{S}(T)$, we consider 100 realizations to obtain the ensemble average of each $\mathcal{S}(T)$.
\\[2mm]
In Fig. \ref{Fig4}(a) we analyze a ring, for this regular network at $T=0$ all the local and non-local measures register $\mathcal{S}(0)=1$, this value occurs when the network has not suffered any damage, therefore it is the largest value that this parameter can reach. With the action of damage at $T=1,2,\ldots,1000$ the ensemble average of all the $\mathcal{S}(T)$ analyzed decrease with $T$. However, in the results for all the realizations, we see that the ensemble average of each measure presents high dispersion, revealing the susceptibility of the ring to damage, the dispersion is huge in the $\mathcal{S}_{\mathrm{MFPT}}(T)$, and $\mathcal{S}_\gamma(T)$ showing how the damage in a single link generates asymmetry that affects the transport on the whole ring. In Fig. \ref{Fig4}(b) we depict the results for a wheel graph. In this case, at $T=0$, the local measure $\mathcal{S}_W(0)=0.5577$ due to an initial bias in the transition matrix $\mathbf{W}$ due to its normalization, an effect that is reduced in non-local measures $\mathcal{S}_{\mathrm{MFPT}}(0)=0.9627$, and $\mathcal{S}_\gamma(0)=0.9755$. We see in the ensemble average the reduction of $\mathcal{S}(T)$ with the damage; but, the values in the realizations are less dispersed in comparison to the results obtained for the ring. This reduced dispersion shows the resistance of the structure at local and global scales, a consequence of the existence of multiple paths that can connect two nodes on the network. On the other hand, the barbell graph analyzed in Fig. \ref{Fig4}(c) combines two fully connected networks with 20 nodes with a linear graph with 10 nodes, each two fully connected subgraphs are resistant to damage due to the diverse alternative routes to defective links. We see that the local measures decaying slowly with $T$ and with low dispersion. However, at a global scale, the nodes that form the linear graph make the structure fragile, this condition of the barbell graph is observed in the non-local measures $\mathcal{S}_{\mathrm{MFPT}}(T)$ and $\mathcal{S}_\gamma(T)$ with fast decay under damage and with high dispersion in the same way as the results observed for the ring in Fig. \ref{Fig4}(a).
\\[2mm]
Regarding the random networks, in Fig. \ref{Fig4}(d) we analyze a Watts-Strogatz network obtained from a regular cyclic structure with four nodes (a ring with two additional links in each node) and a random rewiring with probability $p=0.05$, see Ref. \cite{WattsStrogatz1998}. In this case, the rewiring creates a network with more complexity than the initial structure, the alternative paths connecting two nodes make this structure more resistant to damage. In Fig. \ref{Fig4}(d) it is worth to notice that the average of $\mathcal{S}_W(T)$ and $\mathcal{S}_{\mathrm{MFPT}}(T)$ have a similar behavior showing that the rewiring creates global connectivity of the network, for this reason, $\mathcal{S}_W(T)$ behaves like a non-local measure. In Fig. \ref{Fig4}(e), we explore the evolution of the asymmetry in an Erd\"{o}s-R\'enyi network at the connectivity threshold $p=\log(N)/N$. Due to the reduced number of links the connectivity of the network can be affected with the removal of a link. We see that $\mathcal{S}_{\mathrm{Entropy}}(T)$ decays faster with $T$ in comparison to the results in other networks. For the scale-free network in Fig. \ref{Fig4}(f), $\mathcal{S}_W(0)=0.5166$ showing that the heterogeneity of the network already includes a bias that increases the probability to pass to high connected nodes, the slow variation of $\mathcal{S}_W(T)$ with the damage shows that the distribution of weights in the links has a similar ``complexity'' since the damage is generated with a preferential aggregation. Also, we see a low dispersion of the values $\mathcal{S}_{\mathrm{MFPT}}(T)$, showing that the high number of paths connecting two nodes make this structure capable to resist the damage without creating strong non-local asymmetry. Our findings in Fig. \ref{Fig4} for deterministic and random networks show that the variations of the local asymmetry measures are in a good approximation rescaled versions; in this sense, the two local measures are similar in quantifying variations of the asymmetry due to damage. Something similar occurs with the two non-local measures; however, in this case, the variation of the parameter $\gamma$ can be used to consider other non-local effects.
\\[4mm]
In addition to the temporal evolution, it is important to understand the modifications introduced by the parameter $\alpha$ in Eq. (\ref{OmegaijT}) that modifies the effect of damage in the transport. 
In our `misrepair picture' small $\alpha$ corresponds to high reparation capacity of a damaged link and large $\alpha$ can be related to 
`bad' reparation capacity in the complex system.
In the limit $\alpha=0$, the damage does not alter the transport and $\alpha\to \infty$ is equivalent to the complete removal of the link. In Fig. \ref{Fig5} we analyze the ensemble average of local and non-local symmetry measures $\mathcal{S}$ with $\alpha=0.5,1.0,1.5,2.0$ and $T=1000$ for the networks explored in Fig. \ref{Fig4}. The results are generated with 1000 Monte Carlo simulations of the process, the error bars represent the standard deviation of the data. All the asymmetry measures in the six networks analyzed decrease with the increase of $\alpha$. Regarding the dispersion of the data in the error bars, we see that the values $\mathcal{S}_W(T)$ are less disperse for all the networks; in contrast, the values $\mathcal{S}_\mathrm{MFPT}(T)$ have the largest dispersion revealing that this measure is more susceptible to how damage is distributed. Finally, we see that the non-local measures behave in a similar way with $\alpha$ showing that $\mathcal{S}_\gamma(T)$ with $\gamma=0.5$ is a good measure to quantify the average asymmetry in the transport.
\section{Functionality reduction and aging}
In this section, we study the capacity of the network to perform a specific function and how this property evolves with the accumulation of damage. In the context of transport on networks, we use a `functionality' $\mathcal{F}(T)$ that quantifies the global transport capacity at time $T$ as \cite{Aging_PhysRevE2019}
\begin{equation}\label{F_ratioT}
\mathcal{F}(T)\equiv\frac{\tau(0)}{\tau(T)}
\end{equation}
with
\begin{equation}
\label{globaltime_tau}
\tau(T)= \frac{1}{N}\sum_{j=1}^N \tau_j(T),
\end{equation}
where
\begin{equation}\label{TauiSpect}
\tau_i(T)=\sum_{l=2}^N\frac{1}{1-\lambda_l(T)}\frac{\left\langle i|\phi_l(T)\right\rangle \left\langle\bar{\phi}_l(T)|i\right\rangle}{\left\langle i|\phi_1(T)\right\rangle \left\langle\bar{\phi}_1(T)|i\right\rangle}\, .
\end{equation}
Combining this definition with Eq. (\ref{TijSpect}) and summing over all the initial conditions $i$ considering  weights $P_{i}^{(\infty)}$, we have
\begin{equation}
\sum_{i=1}^N P_{i}^{(\infty)}\left\langle {\cal T}_{ij}(T)\right\rangle = 1+\tau_j(T)
\end{equation}
and, using the mean first return time in Eq. (\ref{meanrcurrencetime}), we have $\tau_j=\sum_{i\neq j}^N P_{i}^{(\infty)}\left\langle {\cal T}_{ij}(T)\right\rangle$, therefore the global time $\tau(T)$ expressed in terms of mean first passage times is given by
\begin{equation}
\tau(T)=\frac{1}{N}\sum_{j=1}^N\sum_{i\neq j}P_{i}^{(\infty)}\left\langle {\cal T}_{ij}(T)\right\rangle.
\end{equation}
In this result, we see that $\tau(T)$ is a global time that gives the weighted average of the number of steps to reach any node of the network. In this way, the definition  $\mathcal{F}(T)$ in Eq. (\ref{F_ratioT}) characterizes globally the effect of the damage suffered
by the whole structure and how evolves the capacity of a
random walker to explore the network. The smaller $\tau(T)$ (i.e. the higher the transport capacity),
the higher the functionality. Since the time $\tau(T) \geq \tau(0)$ in the damaged structure is
greater than in the undamaged structure we have $\mathcal{F}(T) \leq 1$
(equality holds only in the undamaged state) \cite{Aging_PhysRevE2019}.
\\[2mm]
 \begin{figure}[!t] 
 \centering
\includegraphics*[width=0.95\textwidth]{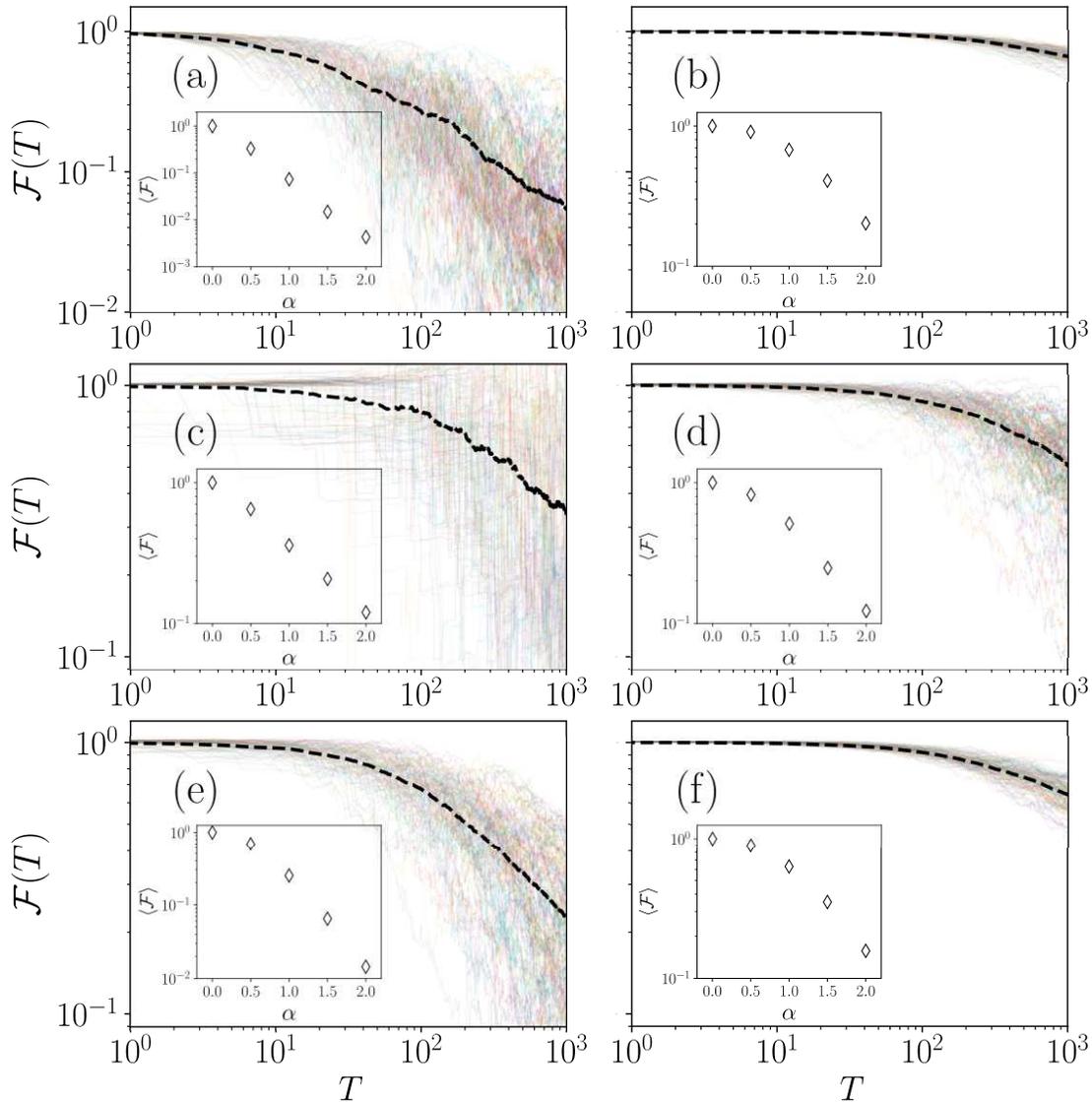}  
\caption{\label{Fig6} Temporal evolution of functionality $\mathcal{F}(T)$ of the transport on networks with cumulative damage with $\alpha=1$. We explore 1000 realizations of this algorithm for: (a) a ring, (b) a wheel graph, (c) a barbell graph, (d) a Watts-Strogatz network, (e) an Erd\"{o}s-R\'enyi and (f) a Barab\'asi-Albert network. Dashed lines represent the ensemble average $\langle\mathcal{F}(T)\rangle$ and the insets depict this quantity for $T=1000$ and different values of $\alpha$.}
 \end{figure}
In Fig. \ref{Fig6}, we analyze the results of Monte Carlo simulation for the evolution of the system subjected to stochastic damage in the edges in the six networks explored before in Figs. \ref{Fig4} and \ref{Fig5}. We analyze the values of $\mathcal{F}(T)$ as a function of $T$ for these structures in different realizations for $\alpha=1$, we see how the global time $\tau(T)$ in Eq. (\ref{globaltime_tau}) differs slightly from the previous value $\tau(T-1)$, i.e. $|\tau(T)-\tau(T-1)|\ll\tau(0)$. Also $\tau(T)-\tau(T-1)$ may be positive or negative since, in particular states, the reduction of the global functionality of a link could produce a small increment of the functionality. However, in general, the most common effect is the damage of the structure, and therefore, we see for $\alpha>0$ that $\mathcal{F}(T)$ starts with $\mathcal{F}(0)=1$ and gradually is reduced with the increase of $T$ in each realization. In Fig. \ref{Fig6}, we also present the average over $1000$ realizations, and from the small deviations observed we can infer that the ensemble average $\langle \mathcal{F}(T)\rangle$ is a good description of the aging in the system, i.e., the global reduction of the functionality.
It is worthy to mention that due to the normalization term in the transition probabilities in Eq. (\ref{transitionPij}); once all the edges have suffered at least one hit, the transition probabilities rescale maintaining the same proportion of damage but increasing the values of $\mathcal{F}(T)$; however, we see that the ensemble average in all the cases decreases with $T$. In the insets in each plot in Fig. \ref{Fig6}, we explore the ensemble average $\langle\mathcal{F}(T)\rangle$ for $T=1000$ as a function of $\alpha$, we see how this parameter modifies the effect of damage at a global scale in the transport.
\\[2mm]
\begin{figure}[!t] 
 \centering
\includegraphics*[width=1.0\textwidth]{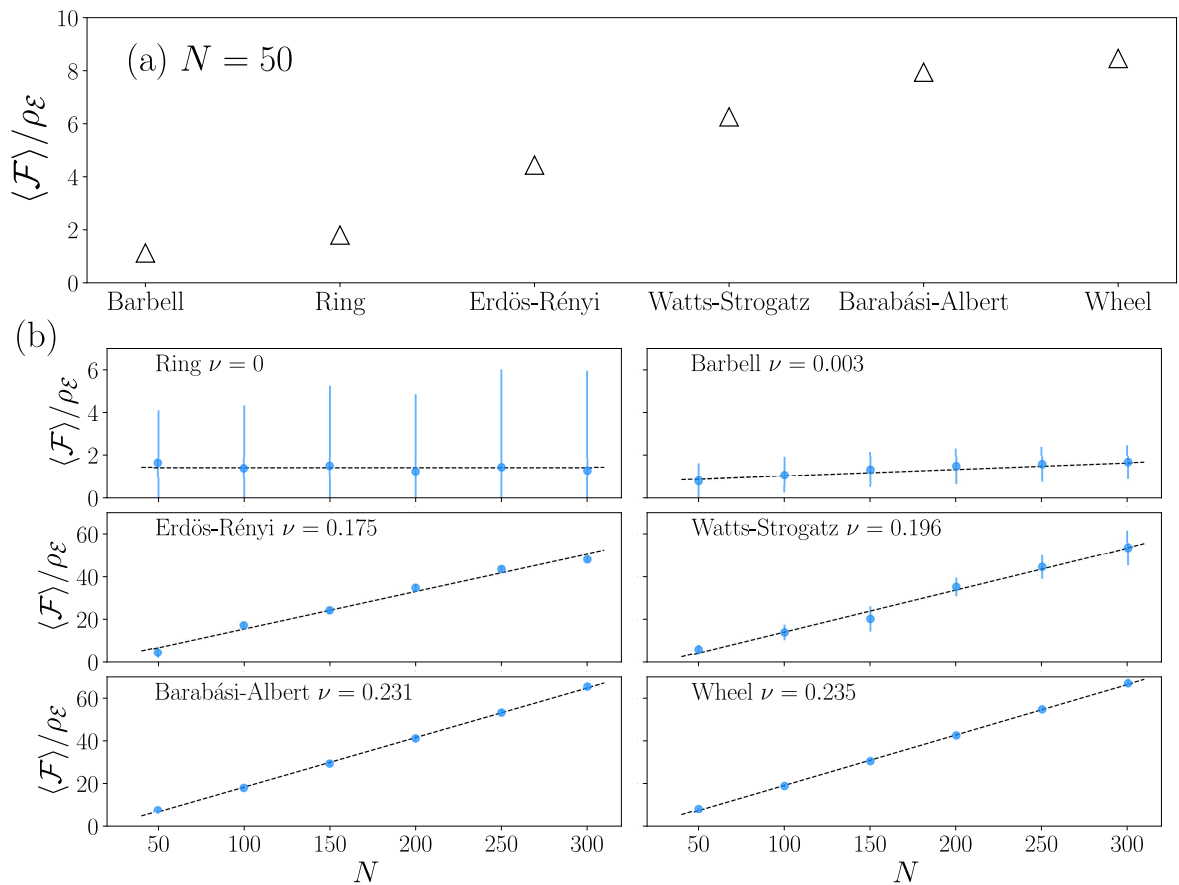}  
\caption{\label{Fig7} Comparison of the values $\langle\mathcal{F}(T)\rangle/\rho_{\mathcal{E}}$ at $T=1000$ for the transport on different network topologies with cumulative damage generated with $\alpha=1$ and 1000 realizations. (a) Networks with $N=50$ in Fig. \ref{Fig2}, (b) $\langle\mathcal{F}\rangle/\rho_{\mathcal{E}}$ as a function of the size $N$, error bars were obtained with the standard deviation of the values. Dashed lines represent the linear fit $\langle\mathcal{F}\rangle/\rho_{\mathcal{E}}=\nu\,N+\kappa$, the values of the slope $\nu$ are presented in each panel.}
\end{figure}
In addition to our discussion about the evolution of asymmetry and the efficiency of transport, the gradual reduction of functionality can be associated with the ability of a system to survive or its ``longevity'' considering its capacity to maintain the global function. We are interested in comparing the previously analyzed systems to determine which topology is the most robust under damage. For this comparison, it is important to notice as a consequence of damage occurring in the edges, more links can generate apparently greater resistance. However, in different cases, these links also mean a cost in the initial configuration of the system. Therefore, it is more convenient to normalize $\langle\mathcal{F}(T)\rangle$ using the quantity
\begin{equation}
\rho_\mathcal{E}=\frac{|\mathcal{E}|}{N(N-1)},
\end{equation}
where $|\mathcal{E}|$ is the total number of edges (including the direction of each line) and $N(N-1)$ is the total number of connections on a fully connected graph without loops.
\\[2mm]
In Fig. \ref{Fig7}(a) we present the values $\langle\mathcal{F}(T)\rangle/\rho_{\mathcal{E}}$ at $T=1000$ for the networks with $N=50$ nodes depicted in Fig. \ref{Fig2} and   analyzed in detail in Figs. \ref{Fig4}--\ref{Fig6}. We sort the structures from the most fragile under damage to the most resistant according to their values $\langle\mathcal{F}(T)\rangle/\rho_{\mathcal{E}}$. We see that the barbell graph ($|\mathcal{E}|=782$) is the graph with the largest number of edges but globally this structure is the most fragile due to the linear path connecting the two fully connected communities, similar resistance to damage is observed for the ring but with a much lower number of links ($|\mathcal{E}|=100$). After the ring, we have the ER network build at the connectivity threshold ($|\mathcal{E}|=138$), a condition that makes the global structure fragile under the removal of a line. These first three structures can be considered simple due to their regularity or completely random character. In contrast, the WS, BA and wheel networks are more complex. Our findings reveal that the WS ($|\mathcal{E}|=200$) produced with rewiring of a regular network has a more robust structure in comparison with the fully random case in the ER network. The more resistant structures are the BA network ($|\mathcal{E}|=194$) and the wheel ($|\mathcal{E}|=196$). A common feature in these two networks is the presence of hubs that increment the connectivity with a high number of routes to connect two nodes on the network and a small average distance between nodes.
\\[2mm]
Finally, in Fig. \ref{Fig7}(b), we analyze the dependence of the values $\langle\mathcal{F}(T)\rangle/\rho_{\mathcal{E}}$ with the size of the network. To this end, we explore networks with $N=100, \,150,\ldots,250,\, 300$ nodes but preserving the same topology of the networks with $N=50$ in Fig. \ref{Fig2}.  Rings and wheels maintain the same characteristics by definition. For the Barbell structures, we consider two fully connected subgraphs with $2 N/5$ nodes connected by a path with $N/5$ nodes. For the WS networks, we maintain the same initial structure and the rewiring probability $p=0.05$, all ER networks are generated at the connectivity threshold, and for BA networks, the algorithm to build the networks is the same. In this manner, each network type maintains the same ``complexity''. Our findings in Fig. \ref{Fig7}(b) reveal the linear relation $\langle\mathcal{F}(T)\rangle/\rho_{\mathcal{E}}=\nu\,N+\kappa$ for $T=1000$ and $\alpha=1$ for each type of network in the interval of values explored. The observed values for the slopes $\nu$ reaffirm that an important factor in aging is the complexity of the structure. We see that the normalized functionality measured by $\langle\mathcal{F}(T)\rangle/\rho_{\mathcal{E}}$ turns out to be much more sensitive to the increase of size in more complex structures.
\section{Conclusions}
In this paper, we explore aging as a consequence of preferential cumulative stochastic damage as a dynamical process on weighted networks. The formalism introduced includes three characteristics: 1) an algorithm to produce preferential random damage on directed links concentrating the damage in particular parts of the structure, 2) the capacity of transport of the structure and, 3) a global measure that quantifies the performance of the structure in a determined configuration. The algorithm for the generation of damage acts on the links producing a bias in the transport
in any type of network that we analyze. 
We use two local measures that include explicitly the transition probabilities between nodes and two non-local measures considering mean first passage times and the fractional Laplacian of a weighted graph, these quantities include information of all the possible paths connecting two nodes on the network.
Finally, we explore aging associated with the reduction of the global transport capacity due to the accumulation of damage. We apply this framework to the study of deterministic networks (ring, wheel, and a barbell) and random networks generated with the Erd\"{o}s-R\'enyi, Watts-Strogatz, and Barab\'asi-Albert algorithms. Our findings allow us to classify the complexity of these structures as a combination of their topology and robustness under damage impact. 
\\[2mm]
The presented methods and approach can be adapted to consider other global measures to characterize the performance of different dynamical processes on networks and provide a framework to understand the relation between the complexity of systems, their fragility, and their lifespan.
\begin{appendix}

\section{Asymptotic fault-distribution - continuous-time limit}
\label{Append1}
In this appendix, we recall the asymptotic fault time-evolution that emerges
from the preferential fault accumulation equation (\ref{problinks}).
To this end, we assume that the network in a time increment $\delta T$ accumulates
the fault measure ${\rm d}\xi \sim \delta T$. 
Then we introduce the total number of faults accumulated up to time ${\cal N}(T,\delta T)=T/\delta t$ 
thus we have from Eq. (\ref{problinks}) 
the relation
\begin{equation}
\label{totaldamage}
\sum_{(k,l) \in \mathcal{E}} (h_{kl}(T)-1) = {\cal N}(T,\delta T)= \frac{T}{\delta T}
\end{equation}
where ${\cal N}(0,\delta T)=0$ reflecting $h_{kl}(0)=1$ where $h_{kl}(T)-1 \in \mathbb{N}_0$ denotes the number of faults
in the edge $(k,l) \in \mathcal{E}$ at time $T$. We observe that the fault numbers in the edges are bounded as $0\leq h_{kl} -1\leq {\cal N}(T,\delta T)$. Let now $M(h,T)$ be the number of edges with $h-1$ faults at time $T$ and 
$P(h,T)=\frac{M(h,T)}{|\mathcal{E}|}$ be the fraction of edges having $h-1$ faults. Then with
$\sum_{h=1}^{\frac{T}{\delta T}+1}M(h,T)=|\mathcal{E}|$ we observe that
\begin{equation}
\label{normalization}
\sum_{h=1}^{\frac{T}{\delta T}+1}P(h,T) =1
\end{equation}
and Eq. (\ref{totaldamage}) can be rewritten as
\begin{equation}
\label{faultnumbers}
\sum_{h=1}^{\frac{T}{\delta T}+1} P(h,T) (h-1) = \frac{T}{\delta T |\mathcal{E}|}.
\end{equation}
Now by introducing the damage measure $\xi(h,\delta T) = (h-1) \delta T \in \delta T \mathbb{N}_0$ with
$P(h,T)= {\cal P}(\xi(h,\delta T),T) \delta T $ where 
we consider the continuous-time limit $\delta T \to {\rm d}\xi \to 0$ 
and assume
$\frac{{\rm d}\xi}{\delta T}=1$. Then with $ \xi(h,\delta T)|_{h=1}=0$ and $ \xi\left(\frac{T}{\delta T}+1 ,\delta T\right)=T$ we arrive at
\begin{equation}
\label{contlim-norm}
\int_0^T {\cal P}(\xi,T){\rm d}\xi =1
\end{equation}
and
\begin{equation}
\label{fault-contilim}
\int_0^T {\cal P}(\xi,T) \xi {\rm d}\xi = \frac{T}{|\mathcal{E}|}.
\end{equation}
Now let us consider the preferential damage accumulation Eq. (\ref{problinks}) which tells us
that a damage arrival ${\rm d}\xi \sim \delta T$ accumulates at with probability $ \sim \xi$. We hence can establish the following master equation 
\begin{equation}
\label{fault-evol-master-eq}
\begin{array}{clc}
\displaystyle   {\cal P}(\xi,T)-{\cal P}(\xi,T-\delta T)  & &\\[2mm] \displaystyle = 
\frac{(\xi -{\rm d}\xi)}{T-\delta T} {\cal P}(\xi-{\rm d}\xi,T-\delta T)- \frac{\xi}{T-\delta T}{\cal P}(\xi,T-\delta T).& &
\end{array}
\end{equation}
By choosing without loss of generality $\frac{\delta T}{d\xi}=1$ and $\delta T \to 0$ this equation writes
\begin{equation}
\label{master-eq}
T \frac{\partial }{\partial T}{\cal P}(\xi,T) = -\frac{\partial }{\partial \xi}\left(\xi{\cal P}(\xi,T)\right).
\end{equation}
In view of normalization (\ref{contlim-norm}) which holds for all $T$ we infer the initial condition
${\cal P}(\xi,0)= \delta(\xi)$ of the form of a Dirac's $\delta$-distribution (concentrated at $0+$). The master equation
(\ref{master-eq}) can be solved by the following separation ansatz
\begin{equation}
\label{separation}
{\cal P}(\xi,T) = U(\xi)V(T)
\end{equation}
to give
\begin{equation}
\label{sepaeq}
\frac{T}{V(T)} \frac{\partial }{\partial T} V(T) =  -\frac{1}{U(\xi)}\frac{\partial }{\partial \xi}\left(\xi U(\xi)\right) = -\lambda
\end{equation}
with the solutions $U(\xi)=C_1\xi^{\lambda-1}$ and $V(T)=C_2 T^{-\lambda}$ where $C_{1,2},\lambda$ are constants to be determined. Hence with $C=C_1C_2$ we can write
\begin{equation}
\label{solution}
{\cal P}(\xi,T)= C \, \frac{\xi^{\lambda-1}}{T^{\lambda}}.
\end{equation}
From the normalization (\ref{contlim-norm}) follows $C=\lambda$. Plugging in this result into Eq. 
(\ref{fault-contilim})
yields $\frac{\lambda}{\lambda +1} = \frac{1}{|\mathcal{E}|}$ thus $\lambda = \frac{1}{|\mathcal{E}|-1}$
and hence 
\begin{equation}
\label{result}
{\cal P}(\xi,T) = \Theta(T-\xi)  \frac{1}{|\mathcal{E}|-1} \frac{\xi^{\frac{1}{|\mathcal{E}|-1}-1}}{T^{\frac{1}{|\mathcal{E}|-1}}}
\end{equation}
where ${\cal P}(\xi,T){\rm d}\xi$ can be conceived as the probability of occurrence of the damage measure within $[\xi, \xi +{\rm d}\xi]$ and in the case of $|\mathcal{E}|\gg 1$ we can replace $|\mathcal{E}|-1 \to |\mathcal{E}|$.
We added in this relation the Heaviside-step function $\Theta(T-\xi)$ to indicate that ${\cal P}(\xi,T)=0$ for $\xi >T$
and we observe that $-1 < \frac{1}{|\mathcal{E}|-1} -1 < 0$ thus the fault accumulation follows a weakly singular power law
$\sim \xi^{\frac{1}{|\mathcal{E}|-1}-1}$ in $\xi$ and the larger $T$ the smaller
the probability to find the a given fixed 
damage value $\xi_0$ in a link which decays with inverse power-law 
$\sim T^{-\frac{1}{|\mathcal{E}|-1}} \to 0$ where this decay is the slower the 
larger the number of edges $\mathcal{E}$. This behavior can be understood as for increasing $T$ 
more and more edges exceed a certain fixed
fault value $\xi_0$ and therefore a fixed value $\xi_0$ is met less likely.
Indeed the self-similar power-law scaling
in the fault distribution (\ref{result}) can be attributed to the emergence of a stochastic fractal distribution in the limit $T/\delta T \to \infty$ as a landmark of complexity \cite{Clauset-et-al_complexity}.

\end{appendix}
\section*{Acknowledgments}
A.P.R. acknowledges support from Ciencia de Frontera 2019 (CONACYT), grant 10782.

\section*{References}
\bibliographystyle{iopart-num}
\providecommand{\noopsort}[1]{}\providecommand{\singleletter}[1]{#1}%
\providecommand{\newblock}{}

\end{document}